\documentstyle[aps,12pt,epsfig]{revtex}
\begin{document}
\title{Chiral symmetry restoration and Medium Effects in 
Relativistic Heavy-Ion Collisions\footnote{Lectures
presented at the CCAST Workshop on Recent Development in
Relativistic Heavy-Ion Collision Physics, Beijing, China,
September 2-6, 1997.}}
\author{G. Q. Li} 
\address{Physics Department, State University of New York at Stony
Brook,\\ Stony Brook, New York 11794, USA}
 
\maketitle
 
\begin{abstract} 

Theoretical and experimental studies of hot and/or dense matter, 
such as is created in high-energy heavy-ion collisions, and 
encountered in compact objects in astrophysics, constitute 
one of the most active frontiers in nuclear physics. 
In these Lectures, we discuss various approaches to 
the description of hot and/or dense matter, including the simple
Skyrme-type parameterization and relativistic Walecka-type models,
as well as microscopic Dirac-Brueckner and QCD sum rule approaches.
As density and/or temperature of the hadronic system increases,
chiral symmetry is gradually restored, as indicated by the 
decrease of quark condensate. This has profound effects 
on the properties of hadrons, especially their masses. We 
review various theoretical predictions for hadron properties 
in dense matter. Experimentally, possible medium modifications 
of hadron properties can be studied through the measurements of particle
spectra, flow, and particularly, electromagnetic observables.
Particle production, especially the production of rare particles
such as kaons, vector mesons, and antiparticles, provides
useful insight into heavy-ion collision dynamics, and 
hadron properties in dense matter. 
Collective flows of various kinds are important observables in
heavy-ion collisions. They probe essentially the entire
reaction process, and thus are very useful for the determination
of the reaction dynamics. They also reflect the properties
of hadrons in dense matter. We discuss flow of nucleons, 
pion, as well as kaons, in heavy-ion collisions from SIS to 
SPS energies (1-200 AGeV). Again, we shall emphasize 
what we can learn about the properties of dense matter and 
in-medium properties of hadrons from the flow study. 
Electromagnetic signals, namely photon and dileptons spectra, 
are considered penetrating probes that may carry undistorted 
information about the early stage of high-energy heavy-ion 
collisions. The observation of the enhancement of low-mass 
dileptons by both the CERES and HELIOS-3 collaborations has 
stimulated a large amount of theoretical activity. We 
discuss various theoretical calculations of dilepton production 
in heavy-ion collisions. We then consider various medium 
effects that have been proposed to explain the enhancement. 
Another piece of experimental data from SPS that have 
been discussed extensively is the single photon spectra 
from the WA80 collaboration. We review various 
hydrodynamics and transport model calculations for direct 
photon production in heavy-ion collisions. 

\end{abstract}

\section{introduction}
 
It is generally accepted that quantum chromodynamics (QCD) is
the ultimate theory of strong interactions \cite{hols92}. At low energy scale
relevant for conventional nuclear physics, QCD exhibits two
important and related features, one is the color confinement and 
the other is approximate chiral symmetry and its spontaneous breaking. 
The latter manifests itself in the smallness of current
quark masses and the non-vanishing quark condensate in vacuum.
The magnitude of the quark condensate, $\langle\bar qq\rangle_0
\approx\langle\bar uu\rangle_0\approx \langle\bar dd\rangle_0$, can 
be estimated from the Gell-Mann$-$Oaks$-$Renner relation, 
\begin{equation}
m_\pi^2f_\pi^2=-2{m_q}\langle\bar qq\rangle_0.
\end{equation}
In the above, $m_\pi\approx 138$ MeV and $f_\pi\approx 93$ MeV 
are the pion mass and decay constant, respectively; 
and $m_q=(m_u+m_d)/2\approx 5.5$ MeV is the average up and down 
quark masses. The quark condensate in vacuum 
thus has a value $\langle\bar qq\rangle_0\approx -(245\,{\rm MeV})^3$. 

\begin{figure}[htb]
\vspace{-2cm}
\epsfig{file=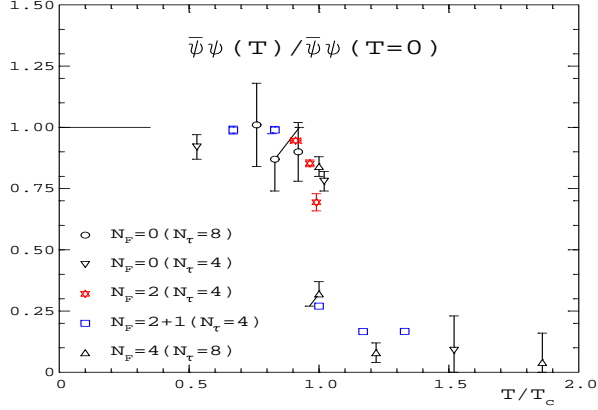,width=5in,height=5in}
\vspace{-4cm}
\caption{Temperature dependence of quark condensate
in the lattice QCD simulation. The different
symbols correspond to different number of flavors and
lattice size. (from
Ref. \protect\cite{karsch95})
\label{qql}}
\end{figure}

As the density and/or temperature of a hadronic system increase, this 
spontaneously broken symmetry is expected to be partially restored, 
so the quark condensate would decrease with increasing density and/or 
temperature. At zero baryon chemical potential, the decrease of quark 
condensate with increasing temperature has been observed in lattice QCD 
simulations \cite{bern92,karsch95,laer96} as well as 
in calculations based on the chiral perturbation theory 
\cite{leut89a,leut89b}. In Fig. \ref{qql} we show the quark condensate 
normalized to that in the vacuum as a function of the normalized
temperature, obtained in lattice QCD simulations with various 
number of flavors and lattice spacing. The quark condensate 
shows weak dependence on the temperature up to about 0.9$T_c$. 
Around the critical tempereture the quark condensate
decreases very rapidly, indicating chiral symmetry restoration. 

\begin{figure}[htb]
\epsfig{file=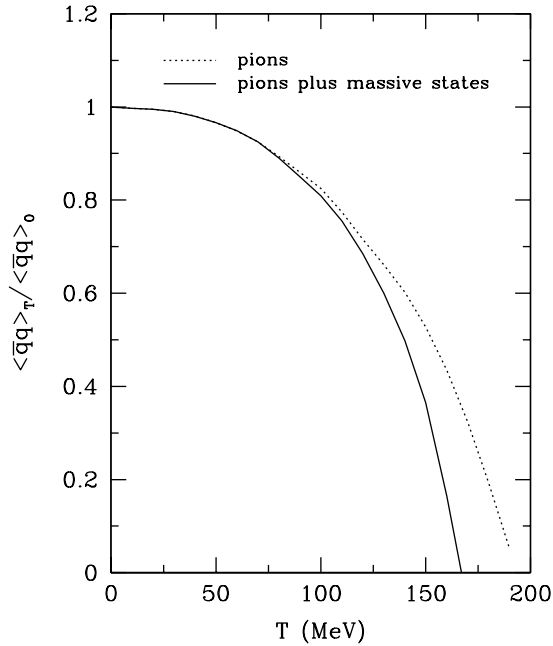,width=4in}
\caption{Temperature dependence of quark condensate
in the chiral perturbation theory. 
The dotted curve includes pions only, while the
solid curve includes massive resonances as well.
(from Ref. \protect\cite{leut89a})
\label{qqc}}
\end{figure}

In the chiral perturbation theory, quark condensate
is found to drop more rapidly at low temperatures than 
that observed in the lattice QCD simulation \cite{leut89a}. 
Expanded in terms of temperature, the leading contribution
to the change of quark condensate is the $T^2$ term,
\begin{eqnarray}
{\langle {\bar q}q\rangle _T\over \langle {\bar q}q\rangle _0}
= \left[1- {n_f^2-1\over n_f} \left({T^2\over 12 f_\pi^2}\right) 
- {n_f^2-1\over 2n_f^2} \left({T^2\over 12f_\pi^2}\right)^2 + O(T^6) 
\right],
\end{eqnarray}
where $n_f$ is the number of flavors.
The results from Ref. \cite{leut89a} 
in the limit of zero quark mass are shown in Fig. \ref{qqc}
as a function of temperature. Again it is seen that the quark 
condensate decreases with increasing temperature, implying
the gradually restoration of the chiral symmetry.
  
\begin{figure}[htb]
\epsfig{file=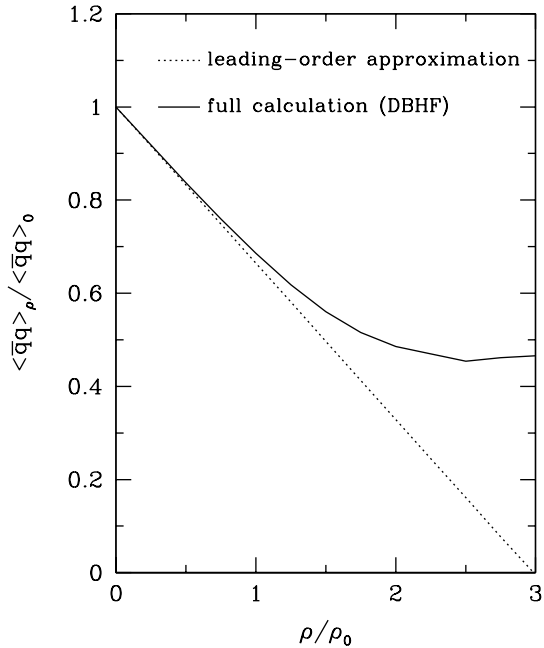,width=4in}
\caption{Density dependence of quark condensate. 
The dotted curve gives the leading order result, while
the solid curves include the higher-order contributions
in the DBHF approach. (from Ref. \protect\cite{liko94a})
\label{qqd}}
\end{figure}

At finite baryon density, model-independent studies 
using the Feynman-Hellmann theorem \cite{lev90,cohen92} have shown that
the ratio of the quark condensate in medium to its value in vacuum
in the leading order in density is given by 
\begin{equation}
\frac{\langle\bar qq\rangle_\rho}{\langle\bar qq\rangle_0}
\approx 1-\frac{\Sigma_{\pi N}}{f_\pi^2m_\pi^2}\rho_N,
\end{equation}
where $\rho_N$ is the nuclear density and $\Sigma_{\pi N}\approx 45$ MeV 
is the ${\pi}N$ sigma term.  At normal nuclear matter density 
$\rho_0\approx 0.16$ fm$^{-3}$, the condensate is seen to decrease 
already by about 1/3.  Higher-order contributions to the quark 
condensate in nuclear matter due to nucleon-nucleon interactions 
have been studied using various models 
\cite{cohen92,birse93,eric93,liko94a,dey95,brock96}.
It is found that at normal nuclear matter density they change the 
leading-order result by only about 5\% \cite{liko94a,brock96}. 
At higher densities, the results, however, depend sensitively on
the model for nuclear matter and on the derivatives of meson
masses to the current quark mass, which are not well
known. We show in Fig. \ref{qqd} the quark condensate 
as a function of density. The dashed curve gives the leading order
results, and solid curve includes higher-order corrections
based on the Dirac-Brueckner-Hartree-Fock (DBHF) calculation 
\cite{liko94a}.
 
\begin{figure}
\vskip 10cm
\caption{QCD phase diagram in the temperature-density plane.
\label{phase}}
\end{figure}

Closely related to the chiral symmetry restoration is the
deconfinement phase transition, namely the transition 
from the non-perturbative hadronic matter at low energy 
densities to the perturbative quark gluon plasma (QGP) at high
energy densities where the color degrees of freedom are no longer
confined. In the lattice QCD simulations, this phase
transition is predicted to occur at a critical energy
density of $\varepsilon \approx 1-3$ GeV/fm$^3$. 
The search for and the study of the properties of 
the QGP is one of the most active fields in nuclear physics \cite{harris96}. 
In Fig. \ref{phase} the QCD phase diagram in the temperature-density
plane is shown. Conventional nuclear physics studies the nuclear
matter around normal nuclear matter density and temperature
close to zero. As the temperature of the system increases,
the nuclear matter may undergo the so-called liquid-gas phase
transtion, and mesons, mainly pions, are produced. The
system can be described as an interacting hadronic gas. At even higher
tempearture, the hadronic system may undergo the deconfinement
phase transition into the QGP. The extremely high temperature and
low net baryon system might correspond to the situation
as encountered in the early universe seconds after the big-bang
explosion. In the other extreme, the nuclear matter can be compressed
to many times the normal nuclear matter density, while the
temperature is kept relatively low. This corresponds
to the situation encountered in the cores of neutron stars. 
In relativistic heavy-ion collisions, ususally both baryon 
density and temperature are increased.

To learn about thermodynamics of hadronic matter and GQP, especially the
so-called equation of state (EoS), is one of the primary motivations
of relativistic heavy-ion physics. The EoS is usually expressed
in terms of the energy density or the pressure of the system
as a function of temperature and/or density. At low temperatures,
the system consists mainly of nucleons, which interact with 
each other through effective nucleon-nucleon interactions.
A simple yet quite useful phenomenological model for the
EoS of the dense nuclear matter is the Skyrme model
\cite{sky56,sky59,brink72,brack85,ligq91}. 
Relativisitc models such as Walecka model and its
various extensions \cite{serot86,serot92,serot97} 
have also been used quite often in
the description of dense matter. More microscopically, the
nuclear matter EoS can also be obtained in the framework of
the DBHF approach \cite{shakin83,mal87,mach89}, 
based on the realistic nucleon-nucleon interaction such as 
the Bonn potential \cite{mach87}.
As the temperatrue of the system increases, mesons are
excited and need to be taken into consideration explicitly.
In the extreme case of high temperature and low net baryon
density, the system can be treated as an interacting meson (chiefly pions)
gas. The EoS of such a system has been studied in both the
chiral perturbation theory and effective meson exchange models.
The EoS of the hot and dense matter can also be obtained by
extending Walecka-type model to finite temperature.
Finally, the EoS can also be simulated on the lattice. We will
discuss the modeling of hot and/or dense matter
in Section 2. 

The study of hadron properties in hot and dense matter is
also a very interesting topic \cite{weis93,birse94,brown96a,kokoch97}.
The question of how the decrease of quark condensate in 
hot and/or dense matter manifects itself in experimental
observables is still under intense debate. One 
possibility is that the decrease of quark condensate in 
medium may lead to reduced hadron masses as shown in 
QCD sum-rule studies \cite{cohen91,hat92a,hat92b}. 
We will review in Section 3 the various theoretical 
prediction for the in-medium properties of hadrons.
We will concentrate on the chiral perturbation theory
for kaon and antikaon in-medium properties, and QCD sum
rule for vector meson properties.
 
The only way to create `macroscopic', strongly interacting systems
at finite temperature and/or density in laboratory is through
collisions of heavy nuclei at high energies. Experiments
carried out at various bombarding energies, ranging from
1 AGeV (BEVALAC, SIS), to 10 AGeV (AGS) 
and 200 AGeV (SPS), have shown that one can indeed 
generate systems of large baryon density but moderate 
temperature (1 AGeV), systems of both high baryon density 
and temperature (10 AGeV), and systems of high temperature 
but relatively small baryon density (200 AGeV). Future 
experiments at RHIC-BNL and LHC-CERN colliders are expected 
to create matter at extremely high temperature
which is essentially baryon free. Therefore, the entire
region of the QCD phase diagram can be investigated
through the variation of the bombarding energy \cite{qm95,qm96,qm97}. 

Unfortunately, the dynamics of heavy-ion 
collisions is very complex, involving a violent initial compression, 
which is then followed by a relatively slow expansion and finally reaches 
the freeze out when particle interactions become unimportant.
The entire reaction typically lasts for about a few tens fm/c. The 
interesting physics of chiral symmetry restoration and
QGP formation can only be studied for a few fm/c during the early part 
of the expansion stage when both the density and temperature of the 
hadronic matter are high. This stage of the collision can in principle 
be probed by detecting the emitted electromagnetic radiation such as the 
real and virtual (dilepton) photons. However, both are not easy to 
measure due to their small rates. Furthermore, they can also be produced 
from initial hard collisions and final hadron decays, and this makes
it difficult to extract the signals from the hot dense matter. What are 
usually measured in heavy-ion experiments are instead the momentum 
distributions of hadrons, such as the nucleon, nuclear clusters, pion, 
kaon, etc., which are mostly ejected from the colliding system at freeze 
out. To infer what have happened in the initial hot dense matter 
from the final hadron phase space distributions requires thus a model that 
can describe the whole collision process. Indeed, various 
transport models have been developed during the past ten years for this 
purpose. We note that a number of good review articles
are available on both relativistic and nonrelativistic transport models 
for heavy-ion collisions at various energies
\cite{stock86,bert88,mal90,mos90,aich91,mos93,peil94,bona95,koli96,geig95,wangx96}.  

One type of transport models that are based on the Walecka-type
$\sigma$-$\omega$ model is particularly suitable for the 
study of hadron properties in dense matter
\cite{ko87,ko88,elze87,blat88,jin90}. In this model,
the effective masses of hadrons are reduced by the scalar potentials
while their energies are shifted by the vector potential.
When the density of the system is high, the hadron masses 
are reduced and converted into the field energy. As the
system expands, the hadrons gradually regain their
masses from the field energy. This model provides
a thermodynamically consistent framework to treat the
change of hadron properties in hot and/or dense matter.
This review will thus discuss mainly the results 
obtained in the relativistic transport models based on
the Walecka-type models.

In addition to a good transport model, we also need to identify
a set of good observables that can be used to study the
hadron in-medium properties. The change of hadron 
masses should affect their production thresholds, and
thus their yields and spectra in heavy-ion collisions.
The threshold effects are expected to be most pronounced
for particle production at the so-called subthreshold energies,
namely the beam energies at which these particles
cannot be produced in free space nucleon-nucleon interactions
simply because of the lack of energy.
Usually, the particle production cross sections near
the thresholds  strongly increase with the available energy,
thus a change of hadron masses will lead to very dramatic
change in the particle production cross sections. In Section 4 
we will discuss particle production in heavy-ion collisions
from SIS to SPS energies, but concentrate on that at 
SIS energies.

Another related, but more subtle observable that is useful
for the study of hadron in-medium properties is the collective
flow of particles and fragments. Up to now mainly three
types of flow have been investigated experimentally and
theoretically, namely, the in-plane flow, out-of-plane flow,
and radial flow. In heavy-ion collisions, dense matter provides
strong mean field potentials for hadrons. When propagating
in these mean field potentials, hadron momenta are modified
which will be reflected in their final momentum spectra,
and therefore in various flow observables which are the 
different projections of the competely momentum distribution.
We will discuss flow observables in Section 5. Again we
will concentate on the results at SIS energies, but
will also mention lower NSCL-MSU energies, and higher
AGS and SPS energies.

Both hadron spectra and flow are subject to strong final-state
interactions. On the other hand, electromagnetic observables,
such as photon and dilepton spectra, are considered
penetrating probes of the early stage of heavy-ion collisions
since they interact with the hadronic environment 
relatively weakly. They might carry undistorted 
information about the densest and hottest phases of the
heavy-ion collisions where chiral symmetry restoration and/or
GQP formation are expected. Moreover, since
dilepton mass spectra reflected directly vector meson
masses, they are most ideal for the study of the
in-medium masses of vector mesons, especially that of the
rho meson. We will discuss the electromagtic
observables in Section 6.

This series of lectures ends with a summary and outlook in Section 7.


\begin{thebibliography}{9}


\bibitem{hols92} J. F. Donoghue, E. Golowich, and B. R. Holstein, 
{\it Dynamics of standard model}, (Cambridge University Press, 
Cambridge, 1992).
 
\bibitem{bern92} C. Bernard, M.C. Ogilvie, T.A. DeGrand, 
C. Detar, S. Gottlieb, A. Krasnitz, R.L. Sugar, and
D. Toussaunt, Phys. Rev. D {\bf 45} (1992) 3854.

\bibitem{karsch95} F. Karsch, Nucl. Phys. {\bf A590} (1995) 367c.

\bibitem{laer96} E. Laermann, Nucl. Phys. {\bf A610} (1996) 1c.
 
\bibitem{leut89a} P. Gerber and H. Leutwyler, Nucl. Phys. {\bf B321} (1989)
387. 

\bibitem{leut89b} J. Gasser and H. Leutwyler, Phys. Lett. B {\bf 184}
(1989) 83.
  
\bibitem{lev90} E.G. Drukarev and E.M. Levin, Nucl. Phys. {\bf A511}
(1990) 679.
 
\bibitem{cohen92} T.D. Cohen, R.J. Furnstahl, and K. Griegel, Phys.
Rev. C {\bf 45} (1992) 1881.
 
\bibitem{birse93} M. C. Birse and J. A. McGovern, Phys. Lett.
B {\bf 309} (1993) 231.
 
\bibitem{eric93} G. Chanfray and M. Ericson, Nucl. Phys. {\bf A556}
(1993) 427.
 
\bibitem{liko94a} G.Q. Li and C.M. Ko, Phys. Lett. {\bf B338}
(1994) 118.
 
\bibitem{dey95} A. Delfino, J. Dey, M. Dey, and M. Malheiro,
Phys. Lett. B {\bf 363} (1995) 17.
 
\bibitem{brock96} R. Brockmann and W. Weise, Phys. Lett. B 
{\bf 367} (1996) 40.

\bibitem{harris96} J.W. Harris and B. M\"uller, Ann. Rev. Nucl.
Part. Sci. {\bf 46} (1996) 71.

\bibitem{sky56} T.H.R. Skyrme, Phil. Mag. {\bf 1} (1956) 1043.

\bibitem{sky59} T.H.R. Skyrme, Nucl. Phys. {\bf 9} (1959) 615.

\bibitem{brink72} D. Vautherin and D.M. Brink, Phys. Rev.
C {\bf 5} (1972) 626.

\bibitem{brack85} M. Brack, C. Guet, and H.B. Hakasson,
Phys. Rep. {\bf 123} (1985) 275.

\bibitem{ligq91} G.Q. Li, J. Phys. G {\bf 17} (1991) 1

\bibitem{serot86}  B.D. Serot and J.D. Walecka, Adv.  Nucl.
Phys.  {\bf 16} (1986) 1.
 
\bibitem{serot92} B.D. Serot, Rep. Prog. Phys. {\bf 55} (1992) 1855. 
 
\bibitem{serot97}  B.D. Serot and J.D. Walecka, Int. Jour. Mod.
Phys. (1997)
 
\bibitem{shakin83} M.R. Anastasio, L.S. Celenza, W.S. Pong, and 
C.M. Shakin, Phys. Rep. {\bf 100} (1983) 327.
 
\bibitem{mal87} B. ter Haar and R. Malfliet, Phys. Rep. {\bf 149}
(1987) 207.
 
\bibitem{mach89} R. Machleidt, Adv. Nucl. Phys. {\bf 19} (1989) 189.
 
\bibitem{mach87} R. Machleidt, K. Holinde, and Ch. Elster, Phys.
Rep. {\bf 149} (1987) 1.
 
\bibitem{weis93} W. Weise, Nucl. Phys. {\bf A553} (1993) 59c.
 
\bibitem{birse94} M. C. Birse, J. Phys. G {\bf 20} (1994) 1537.
 
\bibitem{brown96a} G. E. Brown and M. Rho, Phys. Rep. {\bf 269} (1996)
333.
 
\bibitem{kokoch97} C.M. Ko, V. Koch, G.Q. Li, Ann. Rev. Nucl. Part.
Sci. {\bf 47} (1997), to be published.

\bibitem{cohen91} T.D. Cohen, R.J. Furnstahl, and D.K. Griegel, 
Phys. Rev. Lett. {\bf 67} (1991) 961.
 
\bibitem{hat92a} T. Hatsuda and S.H. Lee, Phys. Rev. C {\bf 46} (1992) R34.

\bibitem{hat92b} T. Hatsuda,  Nucl. Phys. {\bf A544} (1992) 27c.
 
\bibitem{qm95} Quark Matter'95, Nucl. Phys. {\bf A590} (1995).

\bibitem{qm96} Quark Matter'96, Nucl. Phys. {\bf A610} (1996).

\bibitem{qm97} Quark Matter'97, Nucl. Phys. (1998).
 
\bibitem{stock86} H. St\"ocker and W. Greiner, Phys. Rep. {\bf 137}
(1986) 277.
 
\bibitem{bert88} G.F. Bertsch and S. Das Gupta, Phys. Rep. {\bf 160}
(1988) 189.

\bibitem{mal90}W. Botermans and R. Malfliet, Phys. Rep. {\bf 198}
(1990) 115.
 
\bibitem{mos90}  W. Cassing, V. Metag, U. Mosel, and K. Niita, Phys. 
Rep. {\bf 188} (1990) 363.
 
\bibitem{aich91} J. Aichelin, Phys. Rep. {\bf 202} (1991) 235.

\bibitem{mos93}B. Bl\"attel, V. Koch, and U. Mosel, Rep. Prog. 
Phys. {\bf 55} (1993) 1. 
 
\bibitem{peil94} G. Peilert, H. St\"ocker, and W. Greiner, 
Rep. Prog. Phys. {\bf 57} (1994) 533.
 
\bibitem{bona95}  A. Bonasera, F. Gulminelli, and J. Molitoris,
Phys. Rep. {\bf 243} (1994) 1.

\bibitem{koli96} C.M. Ko and G.Q. Li, J. Phys. G {\bf 22} (1996) 
1673.

\bibitem{geig95} K. Geiger, Phys. Rep. {\bf 258} (1995) 237.

\bibitem{wangx96} X.-N. Wang, Phys. Rep. {\bf 280} (1997) 287.

\bibitem{ko87}  C.M. Ko, Q. Li, and R. Wang, Phys. Rev. Lett. {\bf 59}
(1987) 1084. 

\bibitem{ko88} C.M. Ko, and Q. Li, Phys. Rev. C {\bf 37} (1988) 2270.
 
\bibitem{elze87} H. Th. Elze, M. Gyulassy, D. Vasak, H. Heinz, H. St\"ocker,
and W. Greiner,  Mod. Phys. Lett. {\bf A2} (1987) 451.
 
\bibitem{blat88} B. Bl\"attle, V. Koch, W. Cassing, and U. Mosel,
Phys. Rev. C {\bf 38} (1988) 1767.

\bibitem{jin90} X. Jin, Y. Zhuo, X. Zhang, and M. Sano,
Nucl. Phys. {\bf A506} (1990) 655.

 
\bibitem{gale87a}C. Gale, G. Bertsch, and S. Das Gupta, 
Phys. Rev. C {\bf 35} (1987) 1666.
 
\bibitem{aich87} J. Aichelin, A. Rosenhauer, G. Peilert, H. St\"ocker,
and W. Greiner, Phys. Rev. Lett. {\bf 58} (1987) 1926.

\bibitem{welk88} G.M. Welke, M. Prakash, T.T.S. Kuo, S. Das Gupta,
and C. Gale, Phys. Rev. C {\bf 38} (1988) 2101.

\bibitem{gale90} C. Gale, G.M. Welke, M. Prakash, S.J. Lee,
and S. Das Gupta, Phys. Rev. C {\bf 41} (1990) 1545.

\bibitem{limach93a} G.Q. Li and R. Machleidt,  Phys. Rev. C
{\bf 48} (1993) 2707.

\bibitem{paris} M. Lacombe, B. Loiseau, J.M. Richard, R. Vinh Mau,
J. Cote, P. Pires, and R. de Tourreil, Phys. Rev.  C {\bf 21}
(1980) 861.

\bibitem{bog77} J. Boguta and A.R. Bodmer, Nucl. Phys. {\bf A292}
(1977) 413.
 
\bibitem{rein89} P.-G. Reinhard, Rep. Prog. Phys. {\bf 52} (1989)
257.
 
\bibitem{liko95a} G.Q. Li and C.M. Ko, Phys. Lett. {\bf B349} (1995)
405.
 
\bibitem{gel95} G. Gelmini and B. Ritzi, Phys. Lett. {\bf B357} (1995) 
431.
 
\bibitem{brown96b} G. E. Brown and M. Rho, Nucl. Phys {\bf A596}
(1996) 503.
 
\bibitem{fst95}R.J. Furnstahl, H.B. Tang, and B.D. Serot, Phys. 
Rev. C {\bf 52} (1995) 1368.

\bibitem{fst96} R.J. Furnstahl, B.D. Serot, and H.B. Tang,
Nucl. Phys. {\bf A598} (1996) 539.
 
\bibitem{fst97} R.J. Furnstahl, B.D. Serot, and H.B. Tang,
Nucl. Phys. {\bf A615} (1997) 441.

\bibitem{gui88} P.A.M. Guichon, Phys. Lett. B {\bf 200} (1988) 235.

\bibitem{sait94} K. Saito and A.W. Thomas, Phys. Lett. {\bf B327}
(1994) 9.

\bibitem{sait95} K. Saito and A.W. Thomas,
Phys. Rev. C {\bf 51} (1995) 1757.
 
\bibitem{sait96} K. Saito, K. Tsushima, and A.W. Thomas,
Nucl. Phys. {\bf A609} (1996) 339.

\bibitem{sait97}  K. Saito, K. Tsushima, and A.W. Thomas,
Phys. Rev. C {\bf 55} (1997) 2637.

\bibitem{hama90} S. Hama, B.C. Clark, E.D. Cooper, H.S. Sherif, 
and R.L. Mercer, Phys. Rev. C {\bf 41} (1990) 2737.

\bibitem{coop93} E.D. Cooper, S. Hama, B.C. Clark, and R.L. Mercer, 
Phys. Rev. C {\bf 47} (1993) 297.

\bibitem{jin96a} X. Jin and B.K. Jennings, Phys. Lett.
B {\bf 374} (1996) 13.

\bibitem{jin96b} X. Jin and B.K. Jennings, Phys. Rev. C
{\bf 54} (1996) 1427.

\bibitem{mill96} P.G. Blunden and G.A. Miller, Phys. Rev.
C {\bf 54} (1996) 359.

\bibitem{brown91a} G.E. Brown and M. Rho, Phys. Rev. Lett. {\bf 66}
(1991) 2720.
 
\bibitem{song95} H.Q. Song and R.K. Su, Phys. Lett. B {\bf 358}
(1995) 179.

\bibitem{song96} H.Q. Song and R.K. Su, J. Phys. G {\bf 22} 
(1996) 1025.

\bibitem{mach90} R. Brockmann and R. Machleidt, Phys. Rev. C {\bf 42}
(1990) 1965.
 
\bibitem{limach92} G.Q. Li, R. Machleidt and R. Brockmann, Phys. Rev.
C {\bf 45} (1992) 2782.
 
\bibitem{toki92} R. Brockmann and H. Toki, Phys. Rev. Lett.
C {\bf 42} (1990) 1981.

\bibitem{gmu92} S. Gmuca, Nucl. Phys. {\bf A547} (1992) 447.

\bibitem{limach93b} G.Q. Li, R. Machleidt, Y.Z. Zhuo,
Phys. Rev. C {\bf 48} (1993) 1062.

\bibitem{fuch95} C. Fuchs, H. Lenske, and H.H. Wolter,
Phys. Rev. C {\bf 52} (1995) 3043.

\bibitem{shen97} H. Shen, Y. Sugahara, and H. Toki,
Phys. Rev. C {\bf 55} (1997) 1211.

\bibitem{muth90} H. M\"uther, R. Machleidt, and R. Brockmann,
Phys. Rev. C {\bf 42} (1990) 1981.

\bibitem{hub93} H. Huber, F. Weber, and M.K. Weigel, Phys.
Lett. B {\bf 317} (1993) 485.

\bibitem{eng94} L. Engvik, M. Hjorth-Jensen, E. Osnes,
G. Bao, and E. Ostgaard,  Phys. Rev. Lett. {\bf 73} (1994) 2650.

\bibitem{lee97a} C.-H. Lee, T.T.S. Kuo, G.Q. Li, and
G.E. Brown, Phys. Rev. C, submitted.

\bibitem{tani95} I. Tanihata, Prog. Part. Nucl. Phys. {\bf 35} (1995)
505.

\bibitem{han95} P.G. Hansen, A.S. Jensen, and B. Jonson,
Ann. Rev. Nucl. Part. Sci. {\bf 45} (1995) 591.

\bibitem{sehn97} L. Sehn, C. Fuchs, and A. Faessler,
Phys. Rev. C {\bf 56} (1997) 216.

\bibitem{lee97b} C.-H. Lee, T.T.S. Kuo, G.Q. Li, and
G.E. Brown, Phys. Lett. B, submitted.

\bibitem{brock97} R. Brockmann, in Proceedings of
International Workshop on Hadrons in Dense Matter, 
GSI, Darmstadt, July 2-4, 1997.

\bibitem{sait89} K. Saito, T. Maruyama, and K. Soutome, Phys. Rev. C 
{\bf 40} (1989) 407.
 
\bibitem{sait90} K. Soutome, T. Maruyama, and K. Saito, Nucl. Phys.
{\bf A507} (1990) 731.
 
\bibitem{furn90} R.J. Furnstahl and B.D. Serot, Phys. Rev. C {\bf 41}
(1990) 262.

\bibitem{pdata} Particle Data Group, Phys. Rev. D 54 (1996) 1.

\bibitem{cley97} J. Cleymans, K. Redlich, and D.K. Srivastava,
Phys. Rev. C {\bf 55} (1997) 1431.

\bibitem{hage68} R. Hagedorn and J. Ranft, Nuovo Cim. Suppl. {\bf 6}
(1968) 169.

\bibitem{hage80} R. Hagedorn and J. Rafelski, Phys. Lett.
B {\bf 97} (1980) 136.

\bibitem{satz80} F. Karsch and H. Satz, Phys. Rev. D {\bf 21}
(1980) 1168
 
\bibitem{kapu83} J.I. Kapusta and K.A. Olive, Nucl. Phys.
{\bf A408} (1983) 478.

\bibitem{raju92} R. Venugopalan and M. Prakash, Nucl. Phys.
{\bf A546} (1992) 718.

\bibitem{rapp96} R. Rapp and J. Wambach, Phys. Rev. C
{\bf 53} (1996) 3057.

\bibitem{ris95} D.H. Rischke, Y. Pursun, and J.A. Maruhn,
Nucl. Phys. {\bf 595} (1995) 383.

\bibitem{wong94} C.Y. Wong, {\it Introduction to Relativistic 
Heavy-Ion Collisions}, (World Scientific, Singapore, 1994).

\bibitem{detar95} C.E. DeTar, {\it Quark Gluon Plasma 2},
ed. R. Hwa, (World Scientific, Sigapore, 1995).

\bibitem{bern97} C. Bernard, T. Blum, C. DeTar, S. Gottlieb, 
K. Rummukainen, U.M. Heller, J.E. Hetrick, D. Toussaint,
L. K\"arkk\"ainen, R.L. Sugar, and M. Wingate, 
Phys. Rev. D {\bf 55} (1997) 6861.

\bibitem{hung95} C.M. Hung and E.V. Shuryak, Phys. Rev. Lett.
{\bf 75} (1995) 4003.

\bibitem{braun96} P. Braun-Munzinger and J. Stachel,
Nucl. Phys. {\bf A606} (1996) 320.


\bibitem{njl61} Y. Nambu and G. Jona-Lasinio, Phys. Rev. 
{\bf 122} (1961) 345.

\bibitem{bern87} V. Bernard, U-G. Meissner, and I. zahed, Phys.
Rev. Lett. {\bf 59} (1987) 966.

\bibitem{asa89} M. Asakawa and K. Yazaki, Nucl. Phys. {\bf A504} (1989) 668.

\bibitem{hen90} E. M. Henley and H. M\"uther, Nucl. Phys. {\bf A513} (1990) 667.

\bibitem{hat94a} T. Hatsuda and T. Kunihilo, Phys. Rep. {\bf 247}
(1994) 221.

\bibitem{brown96c} G.E. Brown, M. Buballa, and M. Rho, Nucl. Phys. 
{\bf A609} (1996) 519.

\bibitem{hat90} T. Hatsuda, H. Hogaasen, and M. Prakash, Phys. Rev.
C {\bf 42}, 2212 (1990).

\bibitem{adami91} C. Adami and G.E. Brown, Z. Phys. A {\bf 340}
(1991) 93.
 
\bibitem{furn92} R.J. Furnstahl, D.K. Griegel, and T.D. Cohen,
Phys. Rev. C {\bf 46} (1992) 1507.
 
\bibitem{jin93} X. Jin, T.D. Cohen, R.J. Furnstahl, and D.K. Griegel,
Phys. Rev. C {\bf 47} (1993) 2882.
 
\bibitem{cohen95} T.D. Cohen, R.J. Furnstahl, D.K. Griegel, and X.
Jin, Prog. Part. Nucl. Phys. {\bf 35} (1995) 221.

\bibitem{furn96} R.J. Furnstahl, X. Jin, and D.B. Leinweber,
Phys. Lett. B {\bf 387} (1996) 253.
 
\bibitem{shif79} M.A. Shifman, A. I. Vainshtein, and V. I. Zakharov, Nucl.
Phys. {\bf B147} (1979) 385; 448; 519.
 
\bibitem{ioff81} B.L. Ioff, Nucl. Phys. {\bf B188} (1981) 317.
 
\bibitem{rein85} L.J. Reinders, H.R. Rubinstein, and S. Yazaki, 
Phys. Rep. {\bf 127} (1985) 1.
 
\bibitem{shif92} M.A. Shifman, {\it Vacuum Structure and QCD Sum Rules}
(North Holland, Amsterdam, 1992).
 
\bibitem{hoff95} M. Hoffman, R. Mattiello, H. Sorge, H. St\"ocker,
and W. Greiner, Phys. Rev. C {\bf 51} (1995) 2095.
 
\bibitem{liba95a} B.A. Li and C.M. Ko, Phys. Rev. C {\bf 52}
(1995) 2037.
 
\bibitem{lee85} H. Esbensen and T.-S.H. Lee, Phys. Rev. C {\bf 32}
(1985) 1966.

\bibitem{john92} M.B. Johnson and D.J. Ernst, Ann. Phys. (N.Y.)
{\bf 219} (1992) 266.

\bibitem{chen93} C.M. Chen, D.J. Ernst, and M.B. Johnson, 
Phys. Rev. C {\bf 47} (1993) R9.
 
\bibitem{jin95a} X. Jin, Phys. Rev. C {\bf 51} (1995) 2260.

\bibitem{dov88} D. J. Millener, C. B. Dover, and A. Gal, Phys. Rev.
C {\bf 38} (1988) 2700.

\bibitem{yam88} Y. Yamamoto, H. Bando, and J. Zofka, Prog. Theor. Phys. 
{\bf 80} (1988) 757.

\bibitem{gib95} B.F. Gibson and E.V. Hungerford III, 
Phys. Rep. {\bf 257} (1995) 349.
 
\bibitem{sch92} J. Schaffner, C. Greiner, and H. St\"ocker,
Phys. Rev. C {\bf 46} (1992) 322.
 
\bibitem{speth93} A. Reuber, K. Holinde, and J. Speth, 
Nucl. Phys. {\bf A570} (1994) 543.
 
\bibitem{brock77} R. Brockmann and W. Weise, Phys. Lett. B {\bf 69}
(1977) 167.
 
\bibitem{bouy77} A. Bouyssy, Nucl. Phys. {\bf A290} (1977) 324.
 
\bibitem{rufa90} M. Rufa, J. Schaffner, J. Maruhn, H. St\"ocker,
W. Greiner, and P.G. Reinhard, Phys. Rev. C {\bf 42} (1990) 2469.
 
\bibitem{glen91} N.K. Glendenning and S.A. Moszkowski, Phys. Rev. Lett. 
{\bf 67} (1991) 2424.

\bibitem{glen93} N.K. Glendenning, D. Van-Eiff, M. Haft, and H. Lenske, 
M.K. Weigel, Phys. Rev. C {\bf 48} (1993) 889.
 
\bibitem{jin95b} X. Jin and M. Nielsen, Phys. Rev. C {\bf 51}
(1995) 347.
 
\bibitem{jin94} X. Jin and R. J. Furnstahl, Phys. Rev. C {\bf 49}
(1994) 1190.
 
\bibitem{weise88} T. Ericson and W. Weise, {\it Pions and Nuclei}
(Clarendon Press, Oxford, 1988).
 
\bibitem{brown75} G.E. Brown and W. Weise, Phys. Rep. {\bf 22}
(1975) 279.
 
\bibitem{pand81} B. Friemann, V.P. Pandharipande, and Q.N. Usmani, 
Nucl. Phys. {\bf A372} (1981) 483.
 
\bibitem{kox89} C.M. Ko, L.H. Xia, and P.J. Siemens, Phys. Lett. 
{\bf B231} (1989) 16.
 
\bibitem{brown89a} G.E. Brown, M. Vicent Vacas, and W. Weise, Nucl. Phys.
{\bf A505} (1989) 823.
 
\bibitem{xia94} L.H. Xia, P.J. Siemens, and M. Soyeur, Nucl. Phys. 
{\bf A578} (1994) 493.
 
\bibitem{henn94} P.A. Henning and H. Umezawa, Nucl. Phys. {\bf A571}
(1994) 617.
 
\bibitem{kor95} C.L. Korpa and R. Malfliet, Phys. Rev. C {\bf 52},
(1995) 2756.
 
\bibitem{brown91b} G.E. Brown, V. Koch, and M. Rho, Nucl. Phys. 
{\bf A535} (1991) 701.
 
\bibitem{thor95} V. Thorsson and A. Wirzba, Nucl. Phys. {\bf A589}, 
(1995) 633.
 
\bibitem{del92} J. Delorme, M. Ericson, and T.E.O. Ericson, Phys.
Lett. B {\bf 291} (1992) 379.
 
\bibitem{meis88} V. Bernard and U. G. Meissner, Nucl. Phys. {\bf A489}
(1988) 647.
 
\bibitem{lutz92} M. Lutz, S. Klimt, and W. Weise, Nucl. Phys. {\bf A542}
(1992) 521.
 
\bibitem{song94a} C. Song, Phys. Lett. {\bf B329} (1994) 312.

\bibitem{kap86} D.B. Kaplan and A.E. Nelson, Phys. Lett.
{\bf B175} (1986)57.

\bibitem{kap87} A.E. Nelson and D.B. Kaplan, Phys. Lett. 
{\bf B192} (1987) 193.

\bibitem{brown87} G.E. Brown, K. Kubodera, and
M. Rho, Phys. Lett. {\bf B192} (1987) 272.

\bibitem{wise91} H.D. Politzer and M.B. Wise, 
Phys. Lett. {\bf B273} (1991) 156.

\bibitem{muto92} T. Muto and T. Tatsumi, Phys. Lett.
{\bf B283} (1992) 165.

\bibitem{yabu93} H. Yabu, S. Nakamura,  F. Myhrer, and
K. Kubodera, Phys. Lett. {\bf B315} (1993) 17.

\bibitem{lee94} G.E. Brown, C.-H. Lee, M. Rho, and
V. Thorsson, Nucl. Phys. {\bf A567} (1994) 937.

\bibitem{lutz94} M. Lutz, A. Steiner, and W. Weise,
Nucl. Phys. {\bf A574} (1994) 755.

\bibitem{sch94} J. Schaffner, A. Gal, I.N. Mishistin, H.
St\"ocker, and W. Greiner, Phys. Lett. {\bf B334} (1994) 268.

\bibitem{koch94a} V. Koch, Phys. Lett. B {\bf 337} (1994) 7.
 
\bibitem{maru94a} T. Maruyama, H. Fujii, T. Muto, and T. Tatsumi,
Phys. Lett. B {\bf 337} (1994) 19.
 
\bibitem{knor95} R. Knorren, M. Prakash, and P. J. Ellis,
Phys. Rev. C {\bf 52} (1995) 3470.
 
\bibitem{lee95} C.-H. Lee, G.E. Brown, D.P. Min, and M.
Rho, Nucl. Phys. {\bf A585} (1995) 401.

\bibitem{kai95} N. Kaiser, P.B. Siegel, and  W. Weise,
Nucl. Phys. {\bf A594} (1995) 325.
 
\bibitem{kai96} N. Kaiser, T. Waas, and W. Weise,
Phys. Lett. {\bf B365} (1996) 12.

\bibitem{waas96} T. Waas, N. Kaiser, and W. Weise, Phys. Lett.
{\bf B379} (1996).

\bibitem{lee96a} C.-H. Lee, Phys. Rep. {\bf 275} (1996) 255.

\bibitem{lee96b} C.-H. Lee, D.P. Min, and M. Rho, Nucl. Phys. 
{\bf A602} (1996) 334.
 
\bibitem{maru96} H. Fujii, T. Maruyama, and T. Tatsumi, Nucl. Phys.
{\bf A597} (1996) 645.
 
\bibitem{kai97} N. Kaiser, T. Waas, and W. Weise, Nucl. Phys.
{\bf A612} (1997) 297.

\bibitem{waas97a} T. Waas, M. Rho, and W. Weise, Nucl. Phys. 
{\bf A617} (1997) 449.

\bibitem{gal94} E. Friedman, A. Gal, and C. J. Batty,
Nucl. Phys. {\bf A579} (1994) 518.

\bibitem{speth90a}R. B\"uttgen, K. Holinde, A. M\"uller-Groeling, 
J. Speth, and P. Wyborny, Nucl. Phys. {\bf A506} (1990) 586.

\bibitem{speth90b} A. M\"uller-Groeling, K. Holinde, and
J. Speth, Nucl. Phys. {\bf A513} (1990) 557.

\bibitem{leut91} J. Gasser, H. Leutwyler, and M.E. Sainio,
Phys. Lett. {\bf B253} (1991) 252.

\bibitem{liu95} S.J. Dong and K.F. Liu, Nucl. Phys.
{\bf B42} (Proc. Suppl.) (1995) 322.

\bibitem{fuku95} M. Fukugita, Y. Kuramashi, M. Okawa, and A. Ukawa, 
Phys. Rev. D {\bf 51}, 5319 (1995).

\bibitem{pand95} V.R. Pandharipande, C.J. Pethick, and
V. Thorsson, Phys. Rev. Lett. {\bf 75} (1995) 4567.

\bibitem{martin81} A.D. Martin, Nucl. Phys. {\bf B179} (1981) 33.

\bibitem{toki94} M. Mizoguchi, S. Hirenzaki, and 
H. Toki, Nucl. Phys. {\bf A567} (1994) 893.

\bibitem{lilee97} G.Q. Li, C.-H. Lee, and G.E. Brown, Nucl. Phys.
{\bf A}, to be published.

\bibitem{shu92}  E. Shuryak and V. Thorsson, Nucl. Phys. {\bf A536}
(1992) 739.
 
\bibitem{koch95a} V. Koch, Phys. Lett. {\bf B351} (1995) 29.

\bibitem{dov82} C. Dover and G.E. Walker, Phys. Rep.
{\bf 89} (1982) 1.

\bibitem{adami93} C. Adami and G.E. Brown, Phys. Rep. {\bf 224}
(1993) 1.
 
\bibitem{jean94} H. C. Jean, J. Piekarewicz, and A. G. Williams, Phys.
Rev. C {\bf 49} (1994) 1981.
 
\bibitem{hat94b} H. Shiomo and T. Hatsuda, Phys. Lett. {\bf B334}
(1994) 281.
 
\bibitem{song95a} C.S. Song, P.W. Xia, and C.M. Ko, Phys. Rev. C {\bf 52}
(1995) 408.  
 
\bibitem{asa94} M. Asakawa and C.M. Ko, Nucl. Phys. {\bf A572} 
(1994) 732.

\bibitem{lein95} X. Jin and D.B. Leinweber, Phys. Rev. C {\bf 52}
(1995) 3344.

\bibitem{asa93} M. Asakawa and C.M. Ko, Nucl. Phys. {\bf A560}
(1993) 399.
 
\bibitem{asa92} M. Asakawa, C. M. Ko, P. Levai, and X.J. Qiu,
Phys. Rev. C {\bf 46} (1992) R1159.
 
\bibitem{herr93} M. Herrmann, B.L. Friman, and W. N\"orenberg, Nucl.
Phys. {\bf A560} (1993) 411.
 
\bibitem{weise97} F. Klingl and W. Weise, Nucl. Phys. {\bf A}, to
be published.

\bibitem{coh95} T. Cohen, Phys. Rev. C {\bf 45} (1995) 833. 

\bibitem{hat96} T. Hatsuda, H. Shiomi, and H. Kuwaraba,
Prog. Theor. Phys. {\bf 95} (1996) 1009.

\bibitem{brown89b} G.E. Brown and M. Rho, Phys. Lett. {\bf B222}
(1989) 324.
 
\bibitem{brown88} G.E. Brown, C.B. Dover, P.B. Siegel, and W. Weise,
Phys. Rev. Lett. {\bf 60} (1988) 2723.
 
\bibitem{chen92}  C.M. Chen and D.J. Ernst, Phys. Rev. C {\bf 45}
(1992) 2019.
 


\bibitem{cug82} J. Cugnon, D. Kinet, and J. Vandermeulen, 
Nucl. Phys. {\bf A379} (1982) 553.
 
\bibitem{ver82} B.J. VerWest and R.A. Arndt, Phys. Rev. C 
{\bf 25} (1982) 1979.

\bibitem{wolf93} Gy. Wolf, W. Cassing, and U. Mosel,
Nucl. Phys. {\bf A552} (1993) 549.

\bibitem{huber94} S. Huber and J. Aichelin, Nucl. Phys. 
{\bf A573} (1994) 587.

\bibitem{sorge95} H. Sorge,  Phys. Rev C {\bf 52} (1995) 3291.

\bibitem{teis97} S. Teis, W. Cassing, M. Effenberger, A. Hofbach,
U. Mosel, and G. Wolf, Z. Phys. A {\bf 356} (1997) 421. 
 
\bibitem{haar87} B. ter Haar and R. Malfliet, Phys. Rev.
C {\bf 36} (1987) 1611.
 
\bibitem{limach93} G.Q. Li and R. Machleidt, Phys. Rev. C. 
{\bf 48} (1993) 1702.

\bibitem{limach94} G.Q. Li and R. Machleidt, Phys. Rev. C. 
{\bf 49} (1994) 566.
 
\bibitem{alm94} T. Alm, G. R\"opke, and M. Schmit, Phys.
Rev. C {\bf 50} (1994) 31.

\bibitem{alm95} T. Alm, G. R\"opke, W. Bauer, F. Daffin, and 
M. Schmit, Nucl. Phys. {\bf A587} (1995) 815.

\bibitem{sue94} E. Suetomi, N. Kishida, and H. Kadotani,
Phys. Lett. B {\bf 333} (1994) 22.
 
\bibitem{dub96} R.R. Dubey, G.S. Khandelwal, F.A. Cucinotta, and
J.W. Wilson, J. Phys. G {\bf 22} (1996) 387.
 
\bibitem{tana96} E.I. Tanaka, H. Horiuchi, and A. Ono, 
Phys. Rev. C {\bf 54} (1996)
 
\bibitem{mao94} G. Mao, Z. Li, and Y. Zhou, Phys. Rev. C. {\bf 49}
(1994) 3137.
 
\bibitem{mao96} G. Mao, Z. Li, and Y. Zhou, and E. Zhao, Phys. Lett. 
{\bf B378} (1996) 5. 
 
\bibitem{mao97} G. Mao, Z. Li, and Y. Zhuo, Phys. Rev. C {\bf 55} (1997)
792.

\bibitem{hong97} B. Hong {\it et al.,} (FOPI collaboration),
nucl-ex/9707001.

\bibitem{ody88} G. Odyniec {\it et al.,} in {\it Proc. of 
the 8th High Energy Heavy-Ion Study}, ed. J. Harris and G. Woznick, 
LBL Report No. 24580, p. 218 (1988).
 
\bibitem{schw94} O. Schwalb {\it et al.,} Phys. Lett. {\bf B321},
(1994) 20.
 
\bibitem{love91} W.A. Love {\it et al}., Nucl. Phys. {\bf A525}
(1991) 601c.
 
\bibitem{strob88} H. Strobele {\it et al.,} Z. Phys. C {\bf 38}
(1988) 89.
 
\bibitem{lee89} K.S. Lee and U. Heinz, Z. Phys. C {\bf 43}
(1989) 425.
 
\bibitem{soll90} J. Sollfrank, P. Koch, and U. Heinz, 
Phys. Lett. B {\bf 252} (1990) 256.
 
\bibitem{liba91a}  B.A. Li and W. Bauer, Phys. Rev. C {\bf 44}
(1991) 450.
 
\bibitem{barz91} H.W. Barz, G. Bertsch, D. Kusnezov, and J. Schulz,
Phys. Lett. B {\bf 254} (1991) 332.
 
\bibitem{brown91c} G.E. Brown, J. Stachel, and G.M. Welke, 
Phys. Lett. B {\bf 253} (1991) 19.
 
\bibitem{ruus90} M. Kataja and P. V. Ruuskanen, Phys. Lett.
B {\bf 243} (1990) 181.
 
\bibitem{yang91} M.I. Gorenstein and S.N. Yang, Phys. Rev. C {\bf 44}
(1991) 2875.
 
\bibitem{xiong93}  L. Xiong, C.M. Ko, and V. Koch, Phys. Rev. C
{\bf 47} (1993) 788.
 
\bibitem{mosel93} W. Ehehalt, W. Cassing, A. Engel, U. Mosel, and Gy. Wolf, 
Phys. Lett. {\bf B298} (1993) 31.
 
\bibitem{shur90} E.V. Shuryak, Phys. Rev. D {\bf 42} (1990) 1764 

\bibitem{shur91} E.V. Shuryak, Nucl. Phys. {\bf A533} (1991) 761.
 
\bibitem{dan95} P. Danilewicz, Phys. Rev. C {\bf 51}, 716 (1995).
 
\bibitem{hel95} J. Helgesson and J. Randrup, Ann. Phys. {\bf 244} 
(1995) 12.
 
\bibitem{hel96}  J. Helgesson and J. Randrup, Nucl. Phys.
{\bf A597} (1996) 672. 

\bibitem{fuch97} C. Fuchs, L. Sehn, E. Lehmann, J. Zipprich,
and A. Faessler, Phys. Rev. C {\bf 55} (1997) 411.

\bibitem{koch93} V. Koch and G.F. Bertsch, Nucl. Phys. {\bf A552}
(1993) 591.
 
\bibitem{pelt97} D. Pelte {\it et al.,} (FOPI collaboration),
Z. Phys. A {\bf 357} (1997) 215.

\bibitem{ahle95} F. Videbak {\it et al.,} (E866 collaboration),
Nucl. Phys. {\bf A590} (1995) 249c.

\bibitem{bogg96} H. Boggild {\it et al.,} (NA44 collaboration),
Phys. Lett. B {\bf 372} (1996) 339.

\bibitem{stockx86} R. Stock, Phys. Rep. {\bf 135} (1986) 259.

\bibitem{kauf81} M. Gyulassy and S.K. Kauffmann, Nucl. Phys. {\bf A362}
(1981) 503.

\bibitem{liba95b} B.A. Li, Phys. Lett. B {\bf 346} (1995) 5. 

\bibitem{barz96} H.W. Barz, Phys. Rev. C {\bf 53} (1996) 2536.

\bibitem{uma97} V.S. Uma Maheswari, C. Fuchs, A. Faessler,
L. Sehn, D.S. Kosov, and Z. Wang, nucl-th/9706004.

\bibitem{cdata} A. Baldini {\it et al.,} Total cross sections for reaction of
high energy particles, (Springer-Verlag, Heidelberg, 1988).
 
\bibitem{peter95} W. Peters, U. Mosel, and A. Engel, 
Z. Phys. {\bf A353} (1995) 333.

\bibitem{taps94} F.D. Berg {\it et al.,} Phys. Rev. Lett.
B {\bf 72} (1994) 977.

\bibitem{dls97} R.J. Porter {\it et al.,} 
Phys. Rev. Lett. {\bf 79} (1997) 1229.

\bibitem{koch97} V. Koch, in Proceedings of International 
Workshop on Hadrons in Dense Matter, GSI, Darmstadt, July 2-4, 1997, 

\bibitem{waas97b} T. Waas and W. Weise, Nucl. Phys. {\bf A}, to
be published.

\bibitem{aich85} J. Aichelin and C.M. Ko, Phys. Rev. Lett. {\bf 55}
(1985) 2661.
 
\bibitem{cug84}  J. Cugnon and R. M. Lombard, Nucl. Phys. {\bf A422}
(1984) 635.
 
\bibitem{sch87}B. Sch\"urmann and W. Zwermann, Phys. Lett. {\bf B183}
(1987) 31.
 
\bibitem{wu89} J.Q. Wu and C.M. Ko, Nucl. Phys. {\bf A499}
(1989) 810.
 
\bibitem{xiong90a}L. Xiong, C. M. Ko, and J. Q. Wu, 
Phys. Rev. C {\bf 42} (1990) 2231.
 
\bibitem{bat92} G. Batko, J. Randrup, and T. Vetter, Nucl. Phys. 
{\bf A536} (1992) 786.
 
\bibitem{ligq92} G.Q. Li, S.W. Huang, T. Maruyama, Y. Lotfy, D.T. Khoa 
and A. Faessler, Nucl. Phys. {\bf A537}, 645 (1992).

\bibitem{lang92} A. Lang, W. Cassing, U. Mosel, and
K. Weber, Nucl. Phys. {\bf A541} (1992) 507.

\bibitem{ligq93} G.Q. Li, A. Faessler and S.W. Huang, 
Prog. Part. Nucl. Phys. {\bf 31} (1993) 159.
 
\bibitem{huang93} S.W. Huang, A. Faessler, G.Q. Li, R.K. Puri,
E. Lehmann, M.A. Martin,and D.T. Khoa, Phys. Lett.
{\bf B298} (1993) 41.

\bibitem{fang93}  X.S. Fang, C.M. Ko, and Y.M. Zheng, Nucl. Phys. 
{\bf A556} (1993) 449.
 
\bibitem{maru94} T. Maruyama, W. Cassing, U. Mosel, 
S. Teis, and K. Weber,  Nucl. Phys. A573 (1994) 653.

\bibitem{aich94} C. Hartnack, J. J\"anicke, L. Sehn, H. St\"ocker,
and J. Aichelin, Nucl. Phys. {\bf A580} (1994) 643.

\bibitem{fang94a} X.S. Fang, C.M. Ko, G.Q. Li, and Y.M. Zheng, 
Phys. Rev. C {\bf 49} (1994) R608.

\bibitem{fang94b} X.S. Fang, C.M. Ko, G.Q. Li, and Y.M. Zheng, 
Nucl. Phys. {\bf A575} (1994) 766.
 
\bibitem{liko95b} G. Q. Li and C. M. Ko, Nucl. Phys. {\bf A594}, 439 (1995).
 
\bibitem{kaosa} E. Grosse, Prog. Part. Nucl. Phys. {\bf 30} (1993) 89. 

\bibitem{kaosb} P.Senger {\it et al.,} Nucl. Phys. {\bf A553} (1993) 757c.

\bibitem{kaosc} D. Miskowiec, {\it et al.,} Phys. Rev. Lett. {\bf 72},
(1994) 3650.

\bibitem{kaosd} P. Senger for the KaoS collaboration, 
Heavy Ion Physics, {\bf 4} (1996) 317.

\bibitem{kaose} R. Elmer {\it et al.,} Phys. Rev. Lett.
{\bf 77} (1997) 4884.

\bibitem{kaosf} R. Barth and the KaoS Collaboration, Phys. 
Rev. Lett., {\bf 78} (1997) 4027.

\bibitem{best97} D. Best and FOPI collaboration, nucl-ex/9704005, 
Nucl. Phys. A, to be published.

\bibitem{naga89} S. Nagamiya, {\it et al.,} Phys. Rev. C {\bf 40}
(1989) 640.
 
\bibitem{carr88} J.B. Carroll, Nucl. Phys. {\bf A488} (1988) 203c.

\bibitem{shor89} A. Shor, {\it et al.,} Phys. Rev. Lett. {\bf 63}
(1989) 2192.

\bibitem{tsu94a} K. Tsushima, S.W. Huang, and A. Faessler,
Phys. Lett. {\bf B337} (1994) 245.

\bibitem{tsu94b} K. Tsushima, S.W. Huang, and A. Faessler,
J. Phys. G {\bf 21} (1995) 33.

\bibitem{rk80} J. Randrup and C. M. Ko, Nucl. Phys. {\bf A343} (1980) 519.

\bibitem{rk83} J. Randrup and C. M. Ko, {\bf A411} (1983) 537.

\bibitem{cosy} J.T. Balewski {\it et al.,} Phys. Lett.
{\bf B388} (1996) 959.

\bibitem{cass97a} W. Cassing, E. L. Bratkovskaya, U. Mosel,
S. Teis, and A. Sibirtsev, Nucl. Phys. {\bf A614} (1997) 415.

\bibitem{brat97} E.L. Bratkovskaya, W. Cassing, and
U. Mosel, nucl-th/9703047. 

\bibitem{laget91} J.M. Laget, Phys. Lett. {\bf B259} (1991) 24.

\bibitem{tsu97} K. Tsushima, A. Sibirtsev, and A.W. Thomas,
Phys. Lett. {\bf B390} (1997) 29.

\bibitem{likoc97} G.Q. Li, C.M. Ko, and W.S. Chung, 
to be published.

\bibitem{peierls} R.F. Peierls, Phys. Rev. Lett. {\bf 6} (1961) 641.

\bibitem{muon} I.F. Ginzburg, Nucl. Phys. B (Proc. Suppl.)
{\bf 51A} (1996) 85.

\bibitem{chung97a} W.S. Chung, G.Q. Li, and C.M. Ko,
Phys. Lett. {\bf B401} (1997) 1.

\bibitem{chung97b} W.S. Chung, G.Q. Li, and C.M. Ko, Nucl. Phys. 
to be published.

\bibitem{ko83} C. M. Ko, Phys. Lett. {\bf B120} (1983) 294.
 
\bibitem{barz85} H.W. Barz and H. Iwe, Phys. Lett.
{\bf B153} (1985) 217.

\bibitem{huang92} S.W. Huang, G.Q. Li, T. Maruyama, and A. Faessler,
Nucl. Phys.  {\bf A547} (1992) 653.

\bibitem{liko94b} G. Q. Li, C. M. Ko, and X. S. Fang,
Phys. Lett. B {\bf 329} (1994) 149.
 
\bibitem{sib97} A. Sibirtsev, W. Cassing, and C. M. Ko, 
Z. Phys. {\bf 358} (1997) 101.

\bibitem{zwer84} W. Zwermann and B. Sch\"urmann, Phys. Lett.
{\bf B145} (1984) 315.

\bibitem{cosy11} J.T. Balewski {\it et al.,} Acta Physica
Polonica {\bf 27B} (1996) 2911.

\bibitem{martin70} A.D. Martin and G.G. Ross, Nucl. Phys.
{\bf B16} (1970) 479.

\bibitem{schr93} A. Schr\"oter, {\it et al.,} Nucl. Phys. {\bf A553} (1993)
775c.
 
\bibitem{schr94} A. Schr\"oter, {\it et al.,}, Z. Phys. A {\bf 350}
(1994) 101.
 
\bibitem{cham} O. Chamberlain, {\it et al.,} Nuovo Cimento, {\bf 3}
(1956) 447.
 
\bibitem{elio} T. Elioff, {\it et al.,} Phys. Rev. {\bf 128}
(1962) 869.
 
\bibitem{dorf} D.E. Dorfan, {\it et al.,}
Phys. Rev. Lett. {\bf 14} (1965) 995.
 
\bibitem{jinr} A.A. Baldin {\it et al.,} Nucl. Phys. {\bf A519} (1990) 407c.
 
\bibitem{kek} J. Chiba, {\it et al.,} Nucl. Phys. {\bf A553}
(1993) 771c.
 
\bibitem{dan90} P. Danielewicz, Phys. Rev. C {\bf 42} (1990) 1564.
 
\bibitem{mos91} G. Batko, W. Cassing, U. Mosel, K. Niita, and Gy. Wolf,
Phys. Lett. {\bf B256} (1991) 331.
 
\bibitem{liko94c}  G.Q. Li, C.M. Ko, X.S. Fang, and Y. M. Zheng, Phys. Rev.
C {\bf 49} (1994) 1139.
 
\bibitem{liko94d} G.Q. Li and C.M. Ko, Phys. Rev. C {\bf 50} 
(1994) 1725.
 
\bibitem{teis94} S. Teis, W. Cassing, T. Maruyama, and U. Mosel, 
Phys. Rev. C {\bf 50} (1994) 388.
 
\bibitem{fae94} G. Batko, A. Faessler, S. W. Huang, E. Lehmann, 
and R.K. Puri, J. Phys. G {\bf 20} (1994) 461.
 
\bibitem{spie95} C. Spieles, M. Bleicher, A. Jahns, R. Mattiello, H. Sorge, 
H. St\"ocker, and W. Greiner, Phys. Rev. C {\bf 53} (1996) 2011.
 
\bibitem{kog88} C.M. Ko and X. Ge, Phys. Lett. {\bf B205} (1988) 195.
 

\bibitem{sch74} W. Scheid, H. M\"uller, and W. Greiner,
Phys. Rev. Lett. {\bf 32} (1974) 741.

\bibitem{gust84} H.A. Gustafsson {\it et al.,} Phys. Rev. Lett.
{\bf 52} (1984) 1590.

\bibitem{ren84} R.E. Renfordt {\it et al.,} Phys. Rev. Lett. {\bf 53}
(1984) 763.

\bibitem{krof89} D. Krofcheck, {\it et al.,} Phys. Rev. Lett.
{\bf 63} (1989) 2028.

\bibitem{ogil89} C.A. Ogilvie {\it et al.,} Phys. Rev. C {\bf 42} (1990)
R10.

\bibitem{sull90} J.P. Sullivan {\it et al.,} Phys. Lett.
B {\bf 249} (1990) 8.

\bibitem{zhang90} W.M. Zhang {\it et al.,} Phys. Rev. C {\bf 42}
(1990) R491.

\bibitem{krof91} D. Krofcheck {\it et al.,} Phys. Rev. C {\bf 43}
(1991) 350.

\bibitem{shen93} W.Q. Shen {\it et al.,} Nucl. Phys. {\bf A551} (1993)
333.

\bibitem{west93} G.D. Westfall {\it et al.,} Phys. Rev. Lett.
{\bf 71} (1993) 1986.

\bibitem{pak96} R. Pak {\it et al.,} Phys. Rev. C {\bf 53} 
(1996) 1469. 

\bibitem{pak97a} R. Pak {\it et al.,} Phys. Rev. Lett. {\bf 78}
(1997) 1022.

\bibitem{pak97b} R. Pak {\it et al.,} Phys. Rev. Lett.
{\bf 78} (1997) 1026.

\bibitem{beav86} D. Beavis {\it et al.,} Phys. Rev. C {\bf 33}
(1986) 1113.

\bibitem{kean88} D. Keane {\it et al.,} Phys. Rev. C {\bf 37} 
(1988) 1447.

\bibitem{gut89}  H. H. Gutbrod, A. M. Poskanzer, and H. G. Ritter, Rep.
Prog. Phys. {\bf 52}, 1267 (1989).
 
\bibitem{gut90} H. H. Gutbrod, K. H. Kampert, B. Kolb, A. M. Poskanzer,
H. G. Ritter, R. Schicker, and H. R. Schmidt, Phys. Rev. C {\bf 42}, 640
(1990).

\bibitem{demo90} M. Demoulins {\it et al.,} Phys. Lett. 
B {\bf 241} (1990) 476.

\bibitem{wang91} S. Wang {\it et al.,} Phys. Rev. C {\bf 44} (1991) 1091.

\bibitem{rami95} V. Ramillien {\it et al.,} (FOPI collaboration)
Nucl. Phys. {\bf A587} (1995) 802.

\bibitem{part95} M.D. Partland {\it et al.,} (EOS collaboration), 
Phys. Rev. Lett. {\bf 75}, 2100 (1995).

\bibitem{lisa95} M.A. Lisa {\it et al.,} (EOS collaboration),
Phys. Rev. Lett. {\bf 75} (1995) 2662.

\bibitem{reis97} W. Reisdorf {\it et al.,} (FOPI collaboration),
Nucl. Phys. {\bf A612} (1997) 493.

\bibitem{chan97} J. Chance {\it et al.,} (EOS collaboration),
Phys. Rev. Lett. {\bf 78} (1997) 2535.

\bibitem{e877a} J. Barrette {\it et al.,} (E877 collaboration),
Phys. Rev. Lett. {\bf 73} (1994) 2532.

\bibitem{e877b} J. Barrette {\it et al.,} (E877 collaboration),
Phys. Rev. C {\bf 51} (1995) 3309.

\bibitem{e877c} J. Barrette {\it et al.,} Nucl. Phys. {\bf A610} 
(1996) 63c.

\bibitem{e877d} J. Barrette {\it et al.,} Phys. Rev. 
C {\bf 55} (1997) 1420.

\bibitem{awes96} T.C. Awes {\it et al.,} (WA80 collaboration),
Phys. Lett. B {\bf 381} (1996) 29.

\bibitem{wien96} T. Wienold {\it et al.,} (NA49 collaboration),
Nucl. Phys. {\bf A610} (1996) 76c.

\bibitem{xu96} L. G. Bearden {\it et al.,} (NA44 collaboration),
Nucl. Phys. {\bf A610} (1996) 175C.

\bibitem{gyu82} M. Gyulassy, K.A. Frankel, and H.
St\"ocker, Phys. Lett. B {\bf 110} (1982) 185.

\bibitem{dan85}  P. Danielewicz and G. Odyniec, Phys. Lett. {\bf B157}
(1985) 146.
 
\bibitem{dan93}  Q.B. Pan and P. Danielewicz, Phys. Rev. Lett. {\bf 70}
(1993) 2062.
 
\bibitem{zhang94} J. Zhang, S. Das Gupta, and C. Gale, 
Phys. Rev. C {\bf 50} (1994) 1617.

\bibitem{had95} F. Haddad {\it et al.,} Phys. Rev. C {\bf 52}
(1995) 2013.

\bibitem{goss89} J. Gosset {\it et al.,} Phys. Rev. Lett.
{\bf 62} (1989) 1251.

\bibitem{kint97} J.C. Kintner {\it et al.,}
Phys. Rev. Lett. {\bf 78} (1997) 4165.

\bibitem{liba91b} B.A. Li, W. Bauer, and G.F. Bertsch, Phys.
Rev. C {\bf 44} (1991) 2095.

\bibitem{bass93} S.A. Bass, C. Hartnack, H. St\"ocker, 
and W. Greiner, Phys. Rev. Lett. {\bf 71} (1993) 1144.

\bibitem{liba94} B.A. Li, Nucl. Phys. {\bf A570} (1994) 797.

\bibitem{bass95} S.A. Bass, C. Hartnack, H. St\"ocker, 
and W. Greiner, Phys. Rev. C {\bf 51} (1995) 3343.

\bibitem{liko95c} G.Q. Li and C.M. Ko, Nucl. Phys. {\bf A594} (1995) 460.

\bibitem{likoli95} G.Q. Li, C.M. Ko, and B.A. Li, Phys. Rev. Lett.
{\bf 74} (1995) 235.

\bibitem{aich96} C. David, C. Hartnack, M. Kerveno, J.-Ch. Le
Pallec, and J. Aichelin, nucl-th/9611016.

\bibitem{wang97} Z.S. Wang, A. Faessler, C. Fuchs, V.S. Uma
Maheswari, and D. Kosov, nucl-th/9706047.

\bibitem{ritm95} J. Ritman {\it et al.,} (FOPI collaboration), Z. Phys. 
{\bf A352} (1995) 355

\bibitem{best96} D. Best, in {\it Advances in Nuclear Dynamics}, 
edited by W. Bauer {\it et al.,} (World Scientific, Singapore, 1996).

\bibitem{herr96} N. Herrmann, Nucl. Phys. {\bf A610} (1996) 49c.

\bibitem{just95} M. Justice {\it et al.,} (EOS
collaboration), Nucl. Phys. {\bf A590} (1995) 549c.
 
\bibitem{liko96a} G.Q. Li and C.M. Ko, Phys. Rev. C {\bf 54}
(1996) 1897.
 
\bibitem{jahns94}  A. Jahns, C. Spieles, H. Sorge, H. St\"ocker, and W.
Greiner, Phys. Rev. Lett. {\bf 72} (1994) 3464.

\bibitem{kaha95} D.E. Kahana, D. Keane, Y. Pang, T. Schlagel, and
S. Wang, Phys. Rev. Lett. {\bf 74} (1995) 4404.

\bibitem{liko96b} G.Q. Li and C.M. Ko, Phys. Rev. C {\bf 54}
(1996) R2159.

\bibitem{hart92} C. Hartnack, M. Berenguer, A. Jahns, A.v. Keitz, 
R. Mattiello, A. Rosenhauer, J. Schaffner, Th. Sch\"onfeld,
H. Sorge, L. Winckelmann, H. St\"ocker, and
W. Greiner, Nucl. Phys. {\bf A538} (1992) 53c.

\bibitem{wang96} S. Wang {\it et al.,} (EOS collaboration),
Phys. Rev. Lett. {\bf 76} (1996) 3911.

\bibitem{brill96} D. Brill {\it et al.,} (KaoS collaboration),
Z. Phys. A {\bf 355} (1996) 61.

\bibitem{lei93} Y. Leifels {\it et al.,} Phys. Rev. Lett. 
{\bf 71} (1993) 963.

\bibitem{vene93} L.B. Venema {\it et al.,} (TAPS
collaboration), Phys. Rev. Lett. {\bf 71} (1993) 835.

\bibitem{sch95} A. Schubert {\it et al.,} Nucl. Phys.
{\bf A583} (1995) 385.

\bibitem{brill93} D. Brill {\it et al.,} (KaoS collaboration),
Phys. Rev. Lett. {\bf 71} (1993) 336.

\bibitem{brill97} D. Brill {\it et al.,} (KaoS collaboration),
Z. Phys. A {\bf 357} (1997) 207.

\bibitem{likob96a} G.Q. Li, C.M. Ko, and G.E. Brown, Phys. Lett. 
{\bf B381} (1996) 17.

\bibitem{siem79} P.J. Siemens and J.O. Rasmussen, Phys. Rev. Lett.
{\bf 42} (1979) 880.

\bibitem{jeon94} S.C. Jeong {\it et al.,} Phys. Rev. Lett.
{\bf 72} (1994) 3468.

\bibitem{pogg95} G. Pogg {\it et al.,} Nucl. Phys.
{\bf A586} (1995) 755.

\bibitem{bond78} J.P. Bondorf, S.I.A. Garpman, and J. Zimanyi,
Nucl. Phys. {\bf A296} (1978) 320.

\bibitem{schn93} E. Schnedermann, J. Sollfrank, and
U. Heinz, Phys. Rev. C {\bf 48} (1993) 2462.

\bibitem{moli85} J.J. Molitoris and H. St\"ocker, Phys.
Rev. C {\bf 32} (1985) 346.

\bibitem{bert87} G.F. Bertsch, W.G. Lynch, and M.B. Tsang,
Phys. Lett. B {\bf 189} (1987) 384.

\bibitem{soff95} S. Soff, S.A. Bass, C. Hartnack, H. St\"ocker,
and W. Greiner, Phys. Rev. C {\bf 51} (1995) 3320.

\bibitem{xu91} H.M. Xu, Phys. Rev. Lett. {\bf 67} (1991) 2769.

\bibitem{liba93} B.A. Li, Phys. Rev. C {\bf 48} (1993) 2415.

\bibitem{ma93} Y.G. Ma W.Q. Shen, J. Feng, and Y.Q. Ma, 
Phys. Rev. C {\bf 48} (1993) R1492.

\bibitem{zhou93} H.B. Zhou, Z.X. Li, and Y.Z. Zhou, Phys. Lett. 
{\bf B318} (1993) 19.
 
\bibitem{zhou94} H.B. Zhou, Z. Li, and Y. Zhuo,
Nucl. Phys. {\bf A580} (1994)627.

\bibitem{zhou95} H.B. Zhou, Z.X. Li, and Y.Z. Zhou, Phys. Rev.
C {\bf 50} (1995) R2664.

\bibitem{ma95} Y.G. Ma and W.Q. Shen, Phys. Rev. C {\bf 51} 
(1995) 3256.

\bibitem{liba96} B.A. Li, Z.Z. Ren, C.M. Ko, and S.J. Yennello,
Phys. Rev. Lett. {\bf 76} (1996) 4492.

\bibitem{liba97} B.A. Li, C.M. Ko, W. Bauer,
nucl-th/9707014.

\bibitem{kaha97} D.E. Kahana, Y. Pang, and E. Shuryak, 
Phys. Rev. C {\bf 56} (1997) 481.

\bibitem{e877e} W.-C Chang, Ph.D. thesis, State University of
New York at Stony Brook, May, 1997.

\bibitem{barr95} J. Barrette {\it et al.,} (E877 collaboration),
Nucl. Phys. {\bf A590} (1995) 259c.

\bibitem{likoli96} B.A. Li, C.M. Ko, and G.Q. Li,
Phys. Rev. C {\bf 54} (1996) 844.

\bibitem{ris96} D.H. Rischke, Nucl. Phys. {\bf A610} (1996) 88c.

\bibitem{wess97} J.P. Wessels, in Proceedings of Hirchegg'97.

\bibitem{sorge97} H. Sorge, Phys. Rev. Lett. {\bf 78} (1997) 2309.

\bibitem{wess95} P. Braun-Munzinger, J. Stachel, J.P. Wessels, and 
N. Xu, Phys. Lett. B {\bf 333} (1995) 33.
 
\bibitem{wess96} P. Braun-Munzinger, J. Stachel, J.P. Wessels, and 
N. Xu, Phys. Lett. B {\bf 365} (1996) 1.
 
\bibitem{sta96} J. Stachel, Nucl. Phys. {\bf A610} (1996) 509c.


\bibitem{fein76} E. Feinberg, Nuovo Cimento, {\bf A34}, 39 (1976).
 
\bibitem{shur78}E.V. Shuryak, Phys. Lett. B {\bf 78}, 150 (1978).

\bibitem{shur80}E.V. Shuryak,  Phys. Rep. {\bf 67}, 71 (1980).
 
\bibitem{chin82}S. A. Chin, Phys. Lett. B {\bf 119}, 51 (1982).
 
\bibitem{domo83} G. Domokos, Phys. Rev. D {\bf 28}, 123 (1983).
 
\bibitem{mc85} L. D. McLerran and T. Tiomela, Phys. Rev. D {\bf 31},
545 (1985).
 
\bibitem{kaj86}K. Kajantie, J. Kapusta, L. McLerran, and A. Mekjian,
Phys. Rev. D {\bf 34}  (1986) 2746.
 
\bibitem{xia89}C. M. Ko and L. H. Xia, Phys. Rev. Lett. {\bf 62}
(1989) 1595.
 
\bibitem{cley91} J. Cleymans, K. Redlich and H. Satz, Z. Phys.
C {\bf 52} (1991) 517.
 
\bibitem{ruu92} P. V. Ruuskanen, Nucl. Phys. {\bf A544} (1992)
169c.

\bibitem{ceres95a} G. Agakichiev {\it et al.}, Phys. Rev. Lett. 
{\bf 75} (1995) 1272.

\bibitem{ceres95b} J. P. Wurm for the CERES Collaboration, Nucl. Phys. 
{\bf A590}  (1995) 103c.

\bibitem{ceres95c} I. Tserruya, Nucl. Phys. {\bf A590} (1995) 127c.

\bibitem{ceres96a} G. Agakichiev {\it et al.,} 
Nucl. Phys. {\bf A610} (1996) 317c. 

\bibitem{ceres96b} A. Drees, Nucl. Phys. {\bf A610} (1996) 536c. 

\bibitem{helios95} M. Masera for the HELIOS-3 Collaboration, Nucl. 
Phys. {\bf A590} (1995) 93c. 

\bibitem{lkb95} G. Q. Li, C. M. Ko, and G. E. Brown, Phys. Rev. Lett.
{\bf 75} (1995) 4007.

\bibitem{lkb96} G. Q. Li, C. M. Ko, and G. E. Brown, 
Nucl. Phys. {\bf A606} (1996) 568.

\bibitem{lkbs96} G. Q. Li, C. M. Ko,
G. E. Brown, and H. Sorge, Nucl. Phys. {\bf A611} (1996) 539. 

\bibitem{cass95} W. Cassing, W. Ehehalt, and C. M. Ko, Phys. Lett. 
{\bf B363} (1995) 35. 

\bibitem{cass96} W. Cassing, W. Ehehalt, and I. Kralik,
Phys. Lett. {\bf B377} (1996) 5. 

\bibitem{gale95} D. K. Srivastava, B. Sinha, and C. Gale, Phys. Rev. 
{\bf C53} (1996) R567. 

\bibitem{wam95} G. Chanfray, R. Rapp, and J. Wambach, Phys. Rev. Lett.
{\bf 76} (1996) 368.

\bibitem{rapp97} R. Rapp, G. Chanfray, and J. Wambach,
Nucl. Phys. {\bf A617} (1997) 472.

\bibitem{koch96} V. Koch and C. Song, Phys. Rev. C {\bf 54} (1996) 1903.

\bibitem{haglin96} K. Haglin, Phys. Rev. C {\bf 53} (1996) R2606.

\bibitem{haglin97} J. Murray, W. Bauer, and K. Haglin, 
hep-ph/9611328.

\bibitem{schul96} H.J. Schulze and D. Blaschke, Phys. Lett.
{\bf B386} (1996) 429.
 
\bibitem{steele96} J.M. Steele, H. Yamagishi, and I. Zahed,
Phys. Lett. {\bf B384} (1996) 255.

\bibitem{steele97} J.M. Steele, H. Yamagishi, and I. Zahed,
hep-ph/9704414.

\bibitem{hung97} C. M. Hung and E. Shuryak, Phys. Rev. C {\bf 56} (1997) 453.

\bibitem{soll97} J. Sollfrank, P. Huovinen, M. Kataja, P.V. 
Ruuskanen, M. Prakash, and R. Venugopalan, Phys. Rev. 
C {\bf 55} (1997) 392.
 
\bibitem{red96} R. Baier, M. Dirks, and K. Redlich,
Phys. Rev. D {\bf 55} (1997) 4344.

\bibitem{friman96} B. Friman, Nucl. Phys. {\bf A610} (1996) 358c.

\bibitem{friman97} B. Friman and H.J. Pirner, 
Nucl. Phys. {\bf A617} (1997) 496.

\bibitem{weise96} F. Klingl and W. Weise, Nucl. Phys. {\bf A606} (1996) 329.

\bibitem{kap96} J. Kapusta, D. Kharzeev, and L. McLerran,
Phys. Rev. D {\bf 53} (1996) 5028.

\bibitem{huang96} Z. Huang and X.-N. Wang,
Phys. Rev. D {\bf 53} (1996) 5041.

\bibitem{song94} C. Song, C. M. Ko, and C. Gale, Phys. Rev. D
{\bf 50} (1994) R1827.
 
\bibitem{gale94} C. Gale and P. Lichard, Phys. Rev. D {\bf 49}
(1994) 3338.
 
\bibitem{haglin95} K. Haglin and C. Gale, Phys. Rev. D {\bf 52}
(1995) 6297.
 
\bibitem{satz86} T. Matsui and H. Satz, Phys. Lett. {\bf B178} (1986) 416.

\bibitem{gavin96} S. Gavin and R. Vogt, Nucl. Phys. {\bf A610} (1996) 442c.

\bibitem{na50} M.C. Abreu {\it et al.} Nucl. Phys. {\bf A610} (1996) 404c.

\bibitem{khar96} D. Kharzeev, Nucl. Phys. {\bf A610} (1996) 418c.

\bibitem{wong96} C.W. Wong, Nucl. Phys. {\bf A610} (1996) 434c

\bibitem{blai96} J.-P. Blaizot and T.-Y. Ollitrault, 
Nucl. Phys. {\bf A610} (1996) 452c.

\bibitem{satz97} H. Satz, hep-ph/9706342

\bibitem{wa80} R. Albrecht {\it et al.,} Phys. Rev. Lett. {\bf 76} (1996) 3506.

\bibitem{wa80a} R. Santo {\it et al.,} Nucl. Phys. {\bf A566} (1994) 61c

\bibitem{wa80b} T.C. Awes {\it et al.,} Nucl. Phys. {\bf A590} (1995) 81c.

\bibitem{ceres96c} R. Baur {\it et al.,} Z. Phys. C {\bf 71} (1996) 577.
 
\bibitem{shur94} E.V. Shuryak and L. Xiong, Phys. Lett. {\bf B333} (1994) 316.

\bibitem{sinha94} D. K. Srivastava and B. Sinha, Phys. Rev. Lett. 
{\bf 73} (1994) 2421.

\bibitem{ornik95} N. Arbex, U. Ornik, M. Pl\"umer, A. Timmermann,
and R.M. Weiner, Phys. Lett. {\bf B345} (1995) 307.

\bibitem{dumi95}A. Dumitru, U. Katscher, J.A. Maruhn, H. St\"ocker,
W. Greiner, and D.H. Rischke,  Phys. Rev. C {\bf 51} (1995) 2166.

\bibitem{fai95} J.J. Neumann, D. Seibert, and G. Fai,
Phys. Rev. C {\bf 51} (1995) 1460.

\bibitem{hira97} T. Hirano, S. Muroya, and M. Namiki,
hep-ph/9612234.

\bibitem{gale97} C. Gale, nucl-th/9706026.

\bibitem{sorge89} H. Sorge, H. St\"ocker, and W. Greiner,
Ann. Phys. {\bf 192} (1989) 266.
 
\bibitem{bert91} P. Danielewicz and G.F. Bertsch, Nucl. Phys.
{\bf A533} (1991) 712.

\bibitem{casse96} W. Ehehalt and W. Cassing, Nucl. Phys. {\bf A602}
(1996) 449.

\bibitem{land85} L. G. Landberg, Phys. Rep. {\bf 128} (1985) 301.
 
\bibitem{bh88} R.K. Bhaduri, {\it Models of Nucleon}, 
(Addison-Wesley, Reading, MA, 1988).

\bibitem{dls88a} G. Roche {\it et al.,} Phys. Rev. Lett. {\bf 61}
(1988) 1069.

\bibitem{dls88b} C Naudet {\it et al.,} Phys. Rev. Lett. {\bf 62}
(1988) 2652. 

\bibitem{drees96} A. Drees, Phys. Lett. {\bf B388} (1996) 380.

\bibitem{gale87} C. Gale and J. Kapusta, Phys. Rev. C {\bf 35} (1987) 2107.

\bibitem{gale88} C. Gale and J. Kapusta, Phys. Rev. C {\bf 38} (1988) 2657.

\bibitem{song95b} C. Song, V. Koch, S.-H. Lee and C.M. Ko,
Phys. Lett. B {\bf 366} (1996) 379.

\bibitem{huang95} Z. Hunag, Phys. Lett. B {\bf 361} (1995) 131. 

\bibitem{lang97} K. Langfeld, H. Reinhardt, and M. Rho,
hep-ph/9703342.

\bibitem{klug97} Y. Kluger, V. Koch, J. Randrup, and
X.-N. Wang, nucl-th/9704018.

\bibitem{cass97} E.L. Bratkovskaya and W. Cassing,
Nucl. Phys. A,  in press.

\bibitem{librown97} G.Q. Li and G.E. Brown, Nucl. Phys. A, submitted.

\bibitem{adam97} G.E. Brown, M.A. Halasz, 
G.Q. Li and J.V. Steele, in preparetion.

\bibitem{sig63} P. Singer, Phys. Rev. {\bf 130} (1963) 2441.

\bibitem{kay84} \"O. Kaymakcalan, S. Rajeev, and J. Schechter,
Phys. Rev. D {\bf 30} (1984) 594.

\bibitem{ziel84} M. Zielinski, Phys. Rev. Lett. {\bf 52} (1984) 1195.

\bibitem{shur92} L. Xiong, E. Shuryak, and G.E. Brown, 
Phys. Rev. D {\bf 46} (1992) 3798.

\bibitem{ros81} J.L. Rosner, Phys. Rev. D {\bf 23} (1981) 1127.

\bibitem{kapu91} J. Kapusta, P. Lichard, and D. Seibert,
Phys. Rev. D {\bf 44} (1991) 2774.

\bibitem{song93} C. Song, Phys. Rev. C {\bf 47} (1993) 2861.

\bibitem{hidd88} M. Bando, T. Kugo, and K. Yamawaki,
Phys. Rep. {\bf 164} (1988) 217.


\end{thebibliography}
\end{document}